\documentclass[twocolumn]{aastex63}

\usepackage{amsmath}
\usepackage[T1]{fontenc}
\usepackage{upgreek}

\newcommand{\koral}{\texttt{Koral}}

\newcommand{\alfven}{Alfv\'{e}n}

\newcommand{\uint}{u_\mathrm{int}}

\newcommand{\mdotedd}{\dot{M}_\mathrm{Edd}}
\newcommand{\Ledd}{L_\mathrm{Edd}}

\newcommand{\mdot}{\dot{M}}

\newcommand{\D}[1]{\;\mathrm{d}#1}

\received{}
\revised{}
\accepted{}
\submitjournal{ApJ Letters}

\shorttitle{Boundary conditions in NS accretion}
\shortauthors{Abarca et al.}
\graphicspath{{./}{figures/}}

\begin{document}

\title{Beamed emission from a neutron-star ULX in a GRRMHD simulation}

\correspondingauthor{David Abarca}
\email{dabarca@camk.edu.pl}

\author[0000-0002-9202-8734]{David Abarca}
\affiliation{Nicolaus Copernicus Astronomical Center of the Polish Academy of Sciences, Bartycka 18, 00-716 Warsaw, Poland}
\author[0000-0001-6173-0099]{Kyle Parfrey}
\affiliation{School of Mathematics, Trinity College Dublin, Dublin 2, Ireland}

\author[0000-0001-9043-8062]{W\l odek Klu\' zniak}
\affiliation{Nicolaus Copernicus Astronomical Center of the Polish Academy of Sciences, Bartycka 18, 00-716 Warsaw, Poland}




\begin{abstract}

We perform a global 2.5D general-relativistic radiation 
magnetohydrodynamic simulation of super-critical accretion onto a 
neutron star with a $2\times 10^{10}$ G dipolar magnetic field, 
as a model of a neutron-star-powered
ultraluminous X-ray source (ULX).
We compute a lower limit on the total luminosity
of $\sim 2.5\,\Ledd$, and find the radiation to be highly
beamed by the accretion disk outflows. The apparent
isotropic luminosity, which is a function of the viewing angle,
reaches a maximum above $100\,\Ledd$, consistent with the
luminosities observed in ULXs.

\end{abstract}



\section{Introduction} \label{sec:intro}
Ultraluminous X-ray sources (ULXs) are extragalactic, non-AGN X-ray 
sources with luminosities exceeding $10^{39}\,\text{erg s}^{-1}$ 
\citep{kaaret+17}. Observations, in a handful of ULXs,
of coherent pulsations of  $\sim 1$  s
periodicities \citep{bachetti+14, furst+16, israel+17a,israel+17b,carpano+18,
motch+14,furst+18,heida+19,trudolyubov+08,
doroshenko+18,tsygankov+17,townsend+17,brightman+18} have shown that at least some of these sources are powered by slowly rotating neutron stars 
accreting above their critical limits,
$\mdotedd = \eta \Ledd/c^2$, where $\eta \sim 0.2$  is the binding 
energy per unit mass at the surface of the neutron
star \citep{ss+86} and $L_\mathrm{Edd} = 4 \pi G M m_p c / \sigma_T$ is the Eddington luminosity of an object with mass, $M$.

For low accretion rates $\eta$ is also the expected radiative
efficiency, and the luminosity is proportional to the mass accretion
rate, $L/L_\mathrm{Edd} = \dot{M}/ \dot{M}_\mathrm{Edd}$.
However, due to their extremely large
optical depths,
accretion disks with $\mdot\gtrsim\mdotedd$ can 
no longer cool efficiently.
The accretion flow traps photons and the advection of radiation becomes the primary mode
of energy transport in the disk \citep{begelman+78, abramowicz+88,sadowski+3d,czerny+19}. The large concentration of photons 
launches a radiation-pressure-driven outflow, which originates at the radius
where the radiation flux through the surface of the disk 
becomes super-Eddington \citep{ss+73}. The outflow 
extends from this radius (referred to as the spherization radius) down
to the inner edge of the accretion disk. 
Because of advection, the value of the spherization radius will differ
somewhat from its classic thin-disk value 
\citep{ss+73}. In fact, a
substantial fraction of the photons in the radiation-pressure dominated
inner disk will be advected to the vicinity of the stellar surface and
released there. How it escapes to infinity is the major focus of this
{\sl Letter}.

\citet{king+01} suggested long before the first pulsating ULXs were observed 
that the outflows from a super-Eddington disk could collimate the emission released near the compact object, in a manner similar to the collimation predicted by the thick disk model of the Warsaw group \citep{abramowicz+78,paczynsky+80}. 
The system would then appear to be very 
bright when viewed face-on, and the inferred isotropic luminosity 
$L_\mathrm{iso} = F/(4 \pi d^2)$, where $F$ is the radiation flux measured by the observer and $d$
is the distance to the source, would be much larger than the true luminosity,
$L$, i.e., the total emitted radiation power. 

An additional interesting feature of pulsating ULXs is their unusually 
high spin-up rates. \citet{kluzniak+15} inferred a dipole field of 
$\sim10^{9}$~G from the spin-up rate of M82 X-2. As more pulsating ULXs 
were found, all with high spin-ups, a model was formed which incorporated
the period, spin-up, and luminosity to predict the magnetic field strength
and intensity of the beaming \citep[][hereafter referred to as the KLK model]{king+17, king+19, king+20}. 
Besides very small values of the beaming factor $b=L/L_\mathrm{iso}$ --- implying a high degree of beaming --- the model
also predicts dipole magnetic fields in the range of 
$10^{9}$--$10^{13}$~G, with most values falling between $10^{10}$--$10^{11}$~G. 

In order to model a neutron star accreting at super-Eddington accretion rates, it is
necessary to run general-relativistic radiative magnetohydrodynamic (GRRMHD) simulations. 
As of writing this letter, there is only one such global simulation 
\citep{takahashi+17} which includes a stellar magnetic field, and two
which do not \citep{takahashi+17b,abarca+18}. 

The simulation discussed in \citet{takahashi+17}, while an important result, 
has some shortcomings. It is unclear how the highly magnetized regions are treated, or what effects the numerical density floor or background atmosphere have on the emerging radiation. The simulation is run for a rather short duration, not allowing adequate time for the outflows to reach a steady state.
To overcome these issues, all of which could potentially contaminate
measurements of the luminosity and flux distribution, we introduce
a scheme which captures the highly magnetized regions of the simulation
more realistically. Such a scheme was introduced by \citet{parfrey+17} and we have
implemented it in the GRRMHD code {\koral}. We ensure the numerical floors do not affect the emerging radiation and run the simulation for a much longer duration.

We wish to investigate the degree to which the radiation produced by an accreting 
magnetized neutron star is beamed. Small values of $b$  and lower neutron-star magnetic fields would support the 
KLK model, while values of $b$ near unity
would indicate that some other configuration must be responsible for such high
observed luminosities. 
Even if we do find the radiation to be highly beamed,
a direct comparison to the KLK model would not be very informative
considering the requirement that the magnetospheric
and spherization radii be quite close to each other. 
A further caveat to consider is that we have aligned the magnetic dipole axis
with the disk axis, so even if we included rotation, 
we would not expect to produce pulsations\footnote{It has even been claimed
that a strong pulsed fraction is inconsistent
with strong beaming \citep{mushtukov+21}.}. 
However, the population of non-pulsed ULXs is 
even larger, and there is no reason to believe
 that some of these may not also be powered by neutron stars.

In Section~\ref{numerics} 
we describe the numerical methods and the simulation setup. 
In Section~\ref{results} we describe the results of the simulation. 
In Section~\ref{postproc} we discuss the effects of beaming and the expected observed
luminosity of the system and in Section~\ref{conclusion}
 we summarize the results and our conclusions.

\section{Numerical Methods}
\label{numerics}

We use the code {\koral} \citep{sadowski+m1,sadowski+dynamo} which solves the conservation equations
of GRRMHD on a static grid in a fixed metric, $g_{\mu\nu}$. 
The evolution equations are given by
\begin{align}
&\nabla_\mu (\rho u^\mu) = 0, \\
&\nabla_\mu T^\mu\phantom{}_\nu = G_\nu, \\
&\nabla_\mu R^\mu\phantom{}_\nu = -G_\nu, \\
&\nabla_\mu F^{* \mu \nu}=0.
\end{align}
The equations correspond to conservation of mass, conservation of total 
energy-momentum with coupling of matter and radiation provided by the 
radiation four-force \citep{mihalas84},
and the source-free Maxwell's equations. Conservation of mass depends on $\rho$, 
the baryon rest-mass density, and $u^\mu$ the gas four-velocity. The stress-energy tensor for a magnetized gas is given by 
\begin{equation}
    T^\mu\phantom{}_\nu = 
    (\rho + p + \uint{} + b^2) u^\mu u_\nu + 
      (p + b^2/2)\delta^\mu_\nu - b^\mu b_\nu \,,
\end{equation}
which makes use of $\uint$, the gas internal energy, $p=(\gamma-1) \uint$
the gas pressure (where $\gamma=5/3$ is the adiabatic index), and
the magnetic field 4-vector
$b^\mu = \frac{1}{2} u_\nu F^{*\mu \nu}$. The spatial 
components of the 
radiation stress-energy tensor, $R^\mu \phantom{}_\nu$, are computed
using the $M_1$ closure scheme \citep{mihalas84, sadowski+m1}.

Written in terms of the Hodge dual of the Faraday tensor, $F^{*\mu\nu}$, 
the source-free Maxwell's equations correspond to the induction equation
(spatial components), and the divergence-free condition of the magnetic field ($t$ component),
both of which are evolved using the flux constrained transport algorithm \citep{toth+00}.

Since we 
study a non-rotating star\footnote{The observed ULX periods
of several seconds make the spin of the neutron star negligible in our
simulations, which typically run for about $\sim 10^5\,GM/c^3$, i.e.,
$\sim0.7$~s of physical time.}  it is sufficient to use the
Schwarszchild metric
with a coordinate system that is logarithmic in radius and stretches from
$r=5\,r_g$ to $r=1000\, r_g$, where $r_g = GM/c^2$ is the 
gravitational radius.
Unless otherwise specified, we adopt units where $G=c=1$. 
Our simulations are run in 2D
axisymmetry with resolution in $r$ and $\theta$ 
corresponding to $[512,510]$ cells.

We initialize the simulation with an equilibrium torus \citep{penna+13} threaded by a
single loop of magnetic field that 
feeds gas to the star at a rate of 
$\sim 20 \mdotedd$. We initialize a stellar dipole field
with a maximum field strength on the stellar surface
of $2 \times 10^{10}\,\text{G}$ using the
potential given in \cite{wasserman+83}. 

Outside the torus, the gas is initialized to a low-density 
background. This creates a large contrast in the magnetic
and rest-mass energy densities. The ratio of these two 
quantities, the magnetization, $\sigma = b^2/(2\rho)$ provides
an indication of where the numerical scheme should start to break
down, with $\sigma \gg 1$ regions being especially prone to error/instabilities.
Our simulation is initialized with a peak magnetization of $\sigma = 10^4$,
and in order to evolve the system we implement the method
described in \citet{parfrey+17}. The basic idea of the scheme
is to divide the gas into contributions from the real GRMHD flow
and from the numerical floor which keeps $\sigma$ from 
becoming too large. When gas is dominated by the numerical floor, 
the density and internal energy are adjusted to their background 
levels and the velocity parallel to the magnetic field, as measured by
the stationary observer, is reduced.  We provide some additional
adjustments that improve the scheme's robustness in the presence of radiation.
We reduce the scattering and absorption opacities of the 
gas dominated by the numerical floor, and we balance
energy gain/loss from round-off errors in the magnetic field
 by respectively subtracting/adding radiation energy during the conserved-to-primitive variable inversion.

We also introduce a new boundary condition that attempts
to mimic the hard surface of the neutron-star crust. We
treat the gas as in \citet{parfrey+17}, allowing it to 
fall through the inner boundary unimpeded. Then, on a cell-by-cell basis, we measure the flux of 
kinetic, thermal, and radiative energy flowing through the boundary
and return a fraction (albedo) of that energy as outflowing radiation. In the calculations reported in this Letter, 
that fraction corresponds to 75\%, but a full study of this boundary
condition for different values of the albedo is underway. 
An important consideration is the actual flux of radiation that 
crosses the inner boundary, which is controlled by the Riemann solver. 
The ghost cells are set to reflect 75\% of 
the inflowing energy \textit{in the ghost cells}. The HLL Riemann solver should pick a value which is roughly halfway between these two\footnote{The left-biased flux is determined by the ghost cells and the right-biased flux by the domain.} fluxes and so in actuality we expect about 12.5\% of the radiation
flux to escape from the domain through the inner boundary.

We run the simulation for $80\,000\, t_g$ where $t_g=GM/c^3$. Normally, in 2D 
axisymmetry, the absence of an MRI-turbulent dynamo leads to decay of the magnetic field. 
This is remedied with the use of a mean-field dynamo
that restores the magnetic field in the accretion disk similarly to how it would be regenerated in 3D \citep{sadowski+dynamo}.

\section{Simulation results}
\label{results}

The field lines of the stellar dipole are deformed to wrap around 
the initial torus, and so far from the star they are out of equilibrium. 
As the simulation starts, the magnetic field quickly
relaxes to a a stable configuration enveloping the tours while the closed
loops near the star are virtually unchanged.

The torus begins to evolve as the magnetorotational instability (MRI) builds up
and the gas begins to accrete. When the gas reaches the stellar magnetic field,
it forces it inward, raising the magnetic pressure until it balances the ram
pressure, at which point the gas begins to slide along magnetic field lines,
forming accretion columns. As the gas hits the inner boundary, it is shocked and 
a large amount of radiation begins to leave the base of the column
perpendicularly through the column's sides. 
As the simulation progresses, the accretion disk converges at progressively 
larger radii 
to its steady state solution, launching outflows that collimate the radiation 
released in the columns and inner parts of the disk by confining it to 
a funnel-like region about the polar axis.

\begin{figure}
    \centering
    \includegraphics[width=\columnwidth]{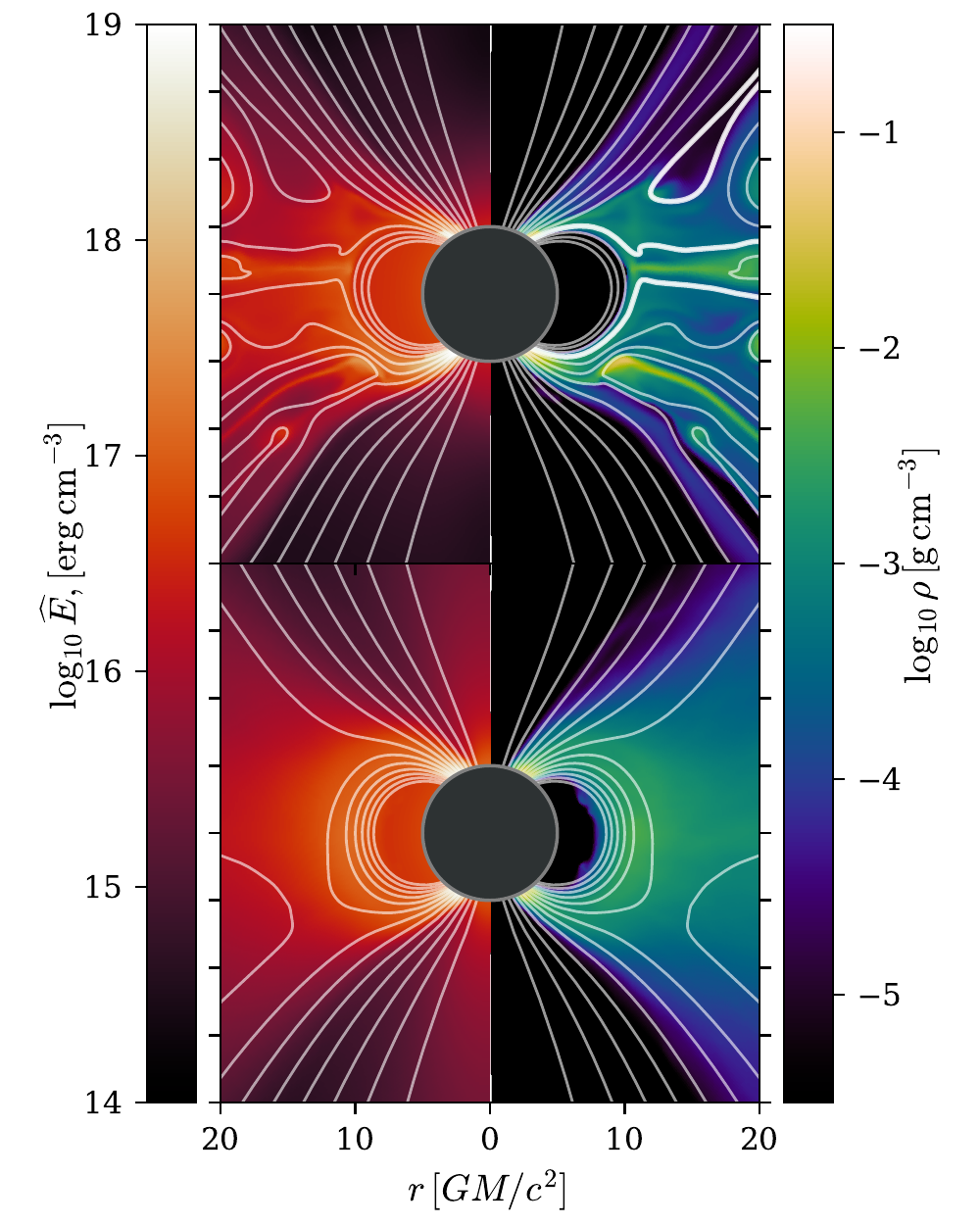}
    \caption{Snapshot (upper) and time-averaged (lower) plots
    of the radiation energy density, left, and the gas rest-mass density, right. We plot
    equispaced contours of $A_\phi$, the $\phi$ component of the vector potential, which correspond to poloidal field lines due to the axisymmetric nature
    of the problem. A bold contour shows the remnant of a torus loop which had just reconnected 
    with the neutron star's dipole.}
    \label{fig:rhorad}
\end{figure}

A snapshot from the simulation at time $t=32\,000\,t_g$ is shown in
the upper panel of Fig.~\ref{fig:rhorad}. The lower panel depicts a time
average from $t=40\,000\,t_g$ to $t=80\,000\,t_g$. The left half of the panel shows
$\widehat{E}$, the radiation energy density in the fluid frame, and
the right half of the panel shows the gas rest-mass density.

The magnetic field of the torus is oriented
to be opposite in direction to the dipole field when they meet, leading 
to reconnection which allows gas to flow smoothly into the accretion 
columns \citep{parfrey+17}. The snapshot shows the remnant of a loop from the torus which
had just reconnected with the stellar dipole in the disk mid-plane, indicated by the bold
contour in the upper panel.

The flow is quite turbulent. The snapshot captures a moment of lower luminosity 
before a high-density parcel of gas below the disk mid-plane enters the column
and collides with the stellar surface, raising the luminosity significantly. 
The long-term effect of successive gas parcels hitting the surface and becoming shocked contributes to the steep radial gradient of radiation energy and gas density at the base of the column. 
This effect is also apparent from the difference in radiation energies in the polar region
between the two panels. It
is also evident that the polar region is largely devoid of gas. The gas is confined
by the magnetic field to mid-latitudes, strongly contrasting to what was observed in \citet{abarca+18},
where the absence of a stellar magnetic field allowed outflowing gas to fill the whole domain.

\begin{figure*}
    \centering
    \includegraphics{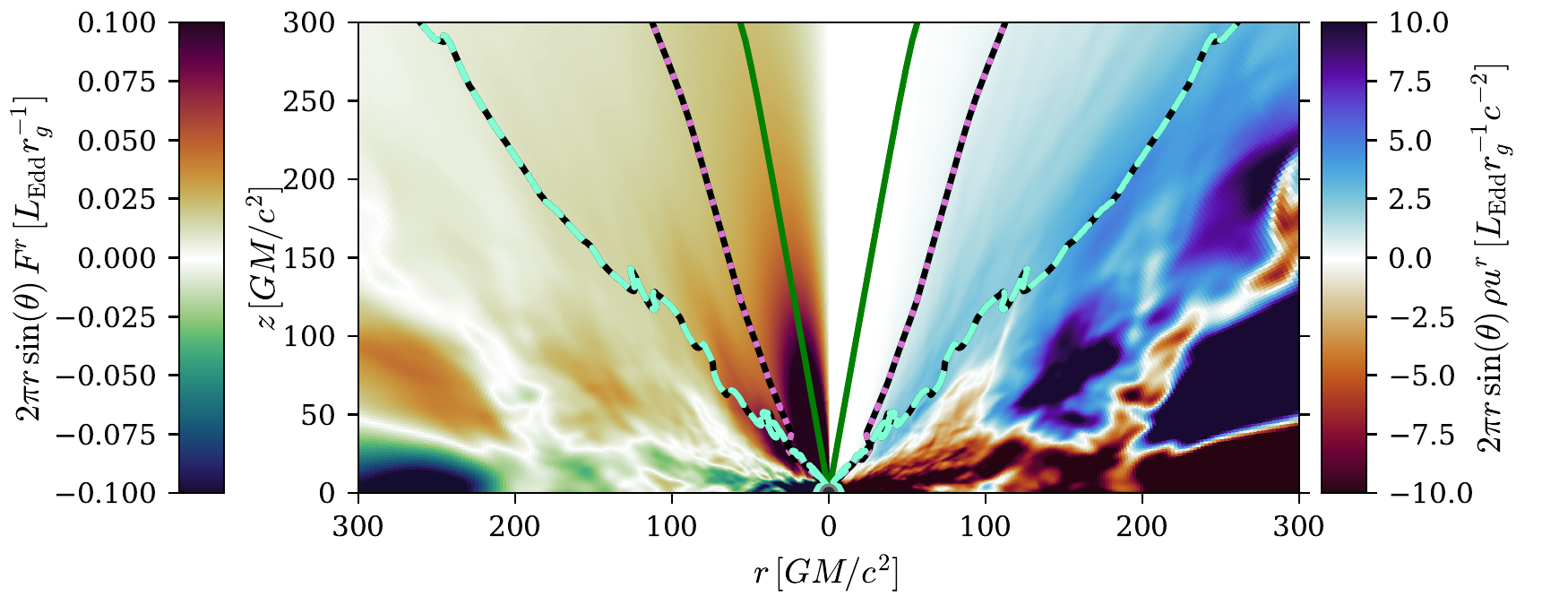}
    \caption{Radial radiative flux, $F^r = R^r_t$ (left), and gas momentum density, $\rho u^r$ (right), integrated into the poloidal plane. Overplotted in solid green is the photosphere at electron scattering depth unity, as measured radially, $\tau_r=1$. The dashed pink and black line shows the photosphere as measured along the $\theta$ direction
    from the axis. The dashed teal and black line shows the zero Bernoulli surface, $Be=0$.}
    \label{fig:outflows}
\end{figure*}

We can study the outflowing gas in more detail by 
considering the azimuthally integrated radial fluxes of gas and radiation, which are shown in 
Fig.~\ref{fig:outflows}. The left half of the panel 
is the quantity $2\uppi r \sin(\theta) F^r$ in units of 
$[\Ledd /r_g$], where $F^r = R^r\phantom{}_t$
is the radial component of the radiation flux 
(or momentum depending on the factor of $c$). One can then integrate by eye over $r\,\mathrm{d}\theta$ to estimate the luminosity. 
In a similar way, the gas momentum is integrated into the poloidal plane giving 
$2\uppi r\sin(\theta)\rho u^r $ in units of $[\Ledd c^{-2}r_g^{-1}$].

Three contours are included. 
The solid green line shows the photosphere
defined by $\tau_r=1$, where $\tau_r$ is the scattering optical depth found by integrating from the outer boundary of the simulation to radius $r$.
We can then assume that all of the 
outflowing radiation above this surface will 
reach the observer giving us a lower limit
on the luminosity if we integrate the flux over the $\theta$ coordinate. 
By definition we can see
very little gas above this surface, and all of the radiation
that is between this line and the axis is expected to reach infinity.
The polar region is completely dominated by a nearly radial flow of
radiation escaping the inner region of the
simulation. 
Surprisingly, a significant fraction of outgoing radiation is excluded by
the $\tau_r=1$ surface. There is very little momentum density in the corresponding
region so one would expect the gas to be optically thin, and for this to be reflected
in the $\tau_r=1$ contour.
However, the outflows do not flow
exactly radially, nor in perfectly straight lines, so the
gas at larger radii is obscuring the inner region. The flux 
at low radii, however, does not know about this gas and freely
streams over a larger range of angles than indicated
by the $\tau_r=1$ surface.  At some point (in our simulation 
between radius 300 and 400 $r_g$)
the radiation scatters off the outflowing gas, becoming 
more confined. This is precisely the radiation-collimation effect 
which should lead to large apparent 
luminosities.

Another consideration is the small density gradient in the
radial direction. This leads to the 
location of the photosphere being quite sensitive to small variations in the 
density and the precise value of the scattering cross section. The draconian
approach of including 100\% of the flux on one side of the contour and excluding
100\% of the flux below the contour may not be appropriate for estimating
the total luminosity, as the photons located immediately
above and below the contour have almost the same probability
of reaching the observer.

We can also measure the optical depth
by integrating along $\theta$ from the axis. 
This surface, $\tau_\theta =1$, is shown in Fig.~\ref{fig:outflows} by the
densely dashed pink and black line. Because
$\tau_\theta$ is more useful for measuring
the amount of radiation which leaves the 
accretion columns (since radiation escapes the column along the
$\theta$ direction), it might provide a more accurate
representation of the radiation which can reach
the observer. The gradient of density
along $\theta$ is much stronger than in the radial 
direction so there is much less uncertainty
in the location of the photosphere in this 
direction. We can see that $\tau_\theta$
is a good indicator for separating the 
very strong radiation flow near the axis from the
less intense radiation flow in the gas outflows.
Also, $\tau_\theta$ is not affected by the 
geometry of the outflow at large radii; however,
the question remains what happens to the radiation
at large radii, or the radiation
which scatters off of the side of the outflow near the accretion column.

The last curve, shown in loosely dashed teal and 
black represents
the surface where the relativistic Bernoulli 
parameter,
\begin{equation}
    Be = - \dfrac{T^t\phantom{}_t + R^t\phantom{}_t + \rho u^t}{\rho u^t},
\end{equation}is equal to zero.  
This surface approximately
splits the domain into energetically
bound and unbound regions.
The gas along this contour would be able to reach 
infinity with zero specific energy if it 
absorbed all of the radiative energy at its location. 
In reality, at some point
the outflow should become diffuse enough that
the radiation escapes. We can therefore use the zero-Bernoulli
surface
to define a region above which we can integrate the radiation flux
to get an upper limit on the luminosity.
If the gas rapidly
becomes optically thin, the luminosity will be 
close to the integral of flux above this surface.
It is more likely that the radiation deposits some
of its momentum into the outflowing gas, lowering the 
luminosity.
One could argue that it is also possible for the gas to cool contributing even more radiation to the total and exceeding this upper limit. However, we can verify that, at least in the simulation domain at radii larger than $\sim 70\,r_g$, the outflows almost exclusively absorb radiation.\footnote{Radiation transfers energy from the hot
inner region to the cooler adiabatically expanding gas in the outer
regions.}

\begin{figure}
    \centering
    \includegraphics{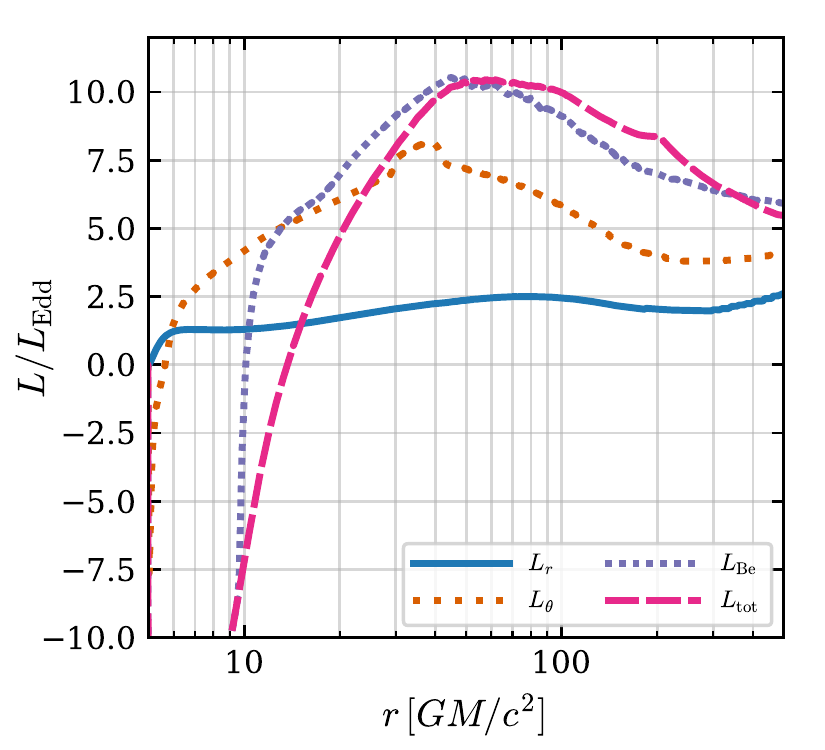}
    \caption{Four measures of the luminosity are plotted as a function of radius.
    In solid blue is $L_r$, loosely dotted orange shows $L_\theta$,
    densely dotted purple is $L_{Be}$, and dashed pink is $L_\mathrm{tot}$.}
    \label{fig:luminosities}
\end{figure}

If we perform the integrals of radiative flux over
spherical shells bounded by these three surfaces, we can plot the luminosity
for each measure as a function of radius as 
shown in Fig.~\ref{fig:luminosities}. In addition
to the three luminosities described above, for reference we 
also plot the total luminosity as integrated
over the entire domain.
Formally we can define the luminosities as follows,
\begin{align}
    L_r &= \int_{\tau_r<1} R^r\phantom{}_t \sqrt{-g} \,\mathrm{d} \theta,\\
    L_\theta &= \int_{\tau_\theta<1} R^r\phantom{}_t \sqrt{-g} \,\mathrm{d} \theta,\\
    L_{Be} &= \int_{Be > 0} R^r\phantom{}_t \sqrt{-g} \,\mathrm{d} \theta,\\   
    L_\mathrm{tot} &= \int_{0}^\pi R^r\phantom{}_t \sqrt{-g} \,\mathrm{d} \theta.
\end{align}

In steady state, the luminosity of a central radiation source would be constant
(apart from redshift factors) with radius if radiative energy were conserved. However, the presence of gas, which can absorb and emit radiation, can change the shape of the luminosity curve even in steady state.
Additionally, because
optical depth is defined along coordinates, and not along the 
path of photons, 
it is possible to arbitrarily 
add or subtract to the luminosity curve if the average path of 
the photons is more complex, such as near the accretion columns.

We have run the
simulation for a sufficiently long duration and taken a 
sufficiently long time average that the accretion disk 
should have reached a steady state out to radius
$60-70\,r_g$ and most
of the turbulence should average away. The disk converges
outwards as the simulation run-time approaches the viscous time at a particular
radius.
The outflows,
which have much larger velocity, converge much faster.
A weak convergence condition for the outflow can be given by
$r/v^r<t_\mathrm{avg}$, where $v^r=u^r/u^t$ is the coordinate velocity,
and $t_\mathrm{avg}$ is the time period over which the simulation data
was averaged. Our data is averaged over a long enough period of 
time that nearly the entirety of the gas outflow is able to satisfy the 
convergence condition. However, one must also take into account the
origin of the outflow. A large portion of the outflowing gas originates
from regions of the disk which have not yet converged, and as we explain
later, this introduces uncertainty into some measures of the
luminosity, especially at larger radii.

Despite all of this, $L_r$ shown as the solid blue line,
is somewhat flat out to
about radius $\sim 500\,r_g$ and so we believe
that $L_r\approx2.5\,\Ledd$ is a good measure
of the total radiative output of the simulation
or at least a suitable lower limit. The steady rise up
to radius $80\,r_g$ is probably due to radiation being emitted in and
emerging from the outflows. Beyond, it is hard to determine
whether the fluctuations are geometrical, or due to 
the unconverged nature of the simulation at large radii.

$L_{Be}$, $L_\mathrm{tot}$, and $L_\theta$, all show
negative values near the star. This is related to the well-understood phenomenon
of photon trapping in super-Eddington accretion disks \citep{sadowski+3d,ohsuga+02}. Most of the radiation
is advected inwards by the optically thick gas before it can 
diffuse out of the disk. 
In our neutron star case some of this energy is released at the surface. It would appear
the photon trapping effect is so strong that even with an albedo of 75\%,
inflowing radiation in the accretion columns dominates energy transport 
near the 
stellar surface. 
$L_\mathrm{tot}$ is also largely dominated by the advection of photons
in the accretion columns and continues to decrease all the way to the surface.

Both $L_r$ and $L_\theta$ rise steeply over the 
first $r_g$ or so above the star due to the radiation
shock. $L_\theta$ continues to rise as
the accretion columns and accretion disk add to
the luminosity. $L_\theta$ includes radiation released from the outflow below radius $\sim70\,r_g$. At larger radii $G_t$ switches sign, and the radiation 
contributing to $L_\theta$ passes through enough 
gas to deposit almost half of its momentum into the outflow
lowering the luminosity to a local minimum of
 $\sim 3.75 L_\mathrm{Edd}$. The location of the $\tau_\theta=1$ surface
 is unaffected by the outer boundary so the steady
rise in $L_\theta$ above $r\sim200\,r_g$
is likely due to the gas becoming steadily thinner, allowing more
of the flux to contribute to the luminosity. 

A similar effect is seen in $L_{Be}$ and
$L_\mathrm{tot}$. They rise sharply
with increasing $r$
up to $r\sim 60\,r_g$, as there is a 
significant amount of radiation advected with the 
outflow,
then drop as momentum starts to be transferred 
to the gas. $L_{Be}$ is integrated over 
regions of the outflow which originate
from parts of the disk which have not yet converged, 
especially beyond radius $r\sim100\,r_g$, which increases the uncertainty in its value, especially at
larger radii. $L_{Be}$ largely follows $L_\mathrm{tot}$, 
although this appears to be largely a coincidence and is due to the equal amounts
of radiation flowing inward and outward over the region where $Be<0$.

\begin{figure}
    \centering
    \includegraphics[width=3.5in]{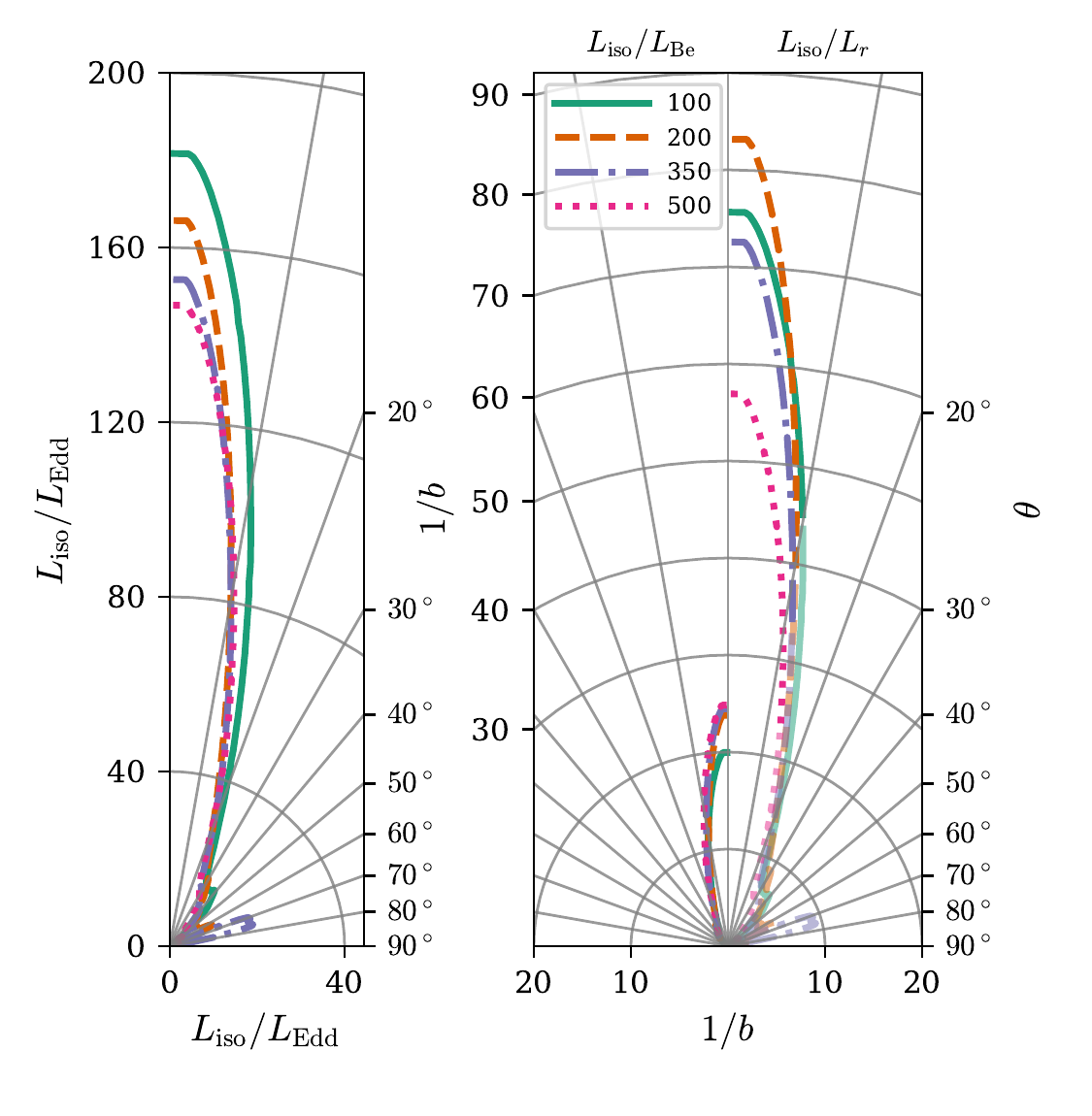}
    \caption{Inferred isotropic flux (left) and beaming pattern (right) as a function of polar angle. For the left panel, the radial coordinate corresponds
    to the isotropic luminosity, $L_\mathrm{iso}=4\pi r^2 F^r$. For the
    right panel, the radial coordinate measures beaming, $1/b$, i.e.,
    the inferred isotropic luminosity a function of angle divided by the total luminosity.
    The left side
    of the right panel is a lower limit for $1/b$ obtained with the total 
    luminosity, $L_\mathrm{Be}$. The right side of the right panel
    assumes the total luminosity is $L$, which provides an upper limit on
    $1/b$. All the quantities are measured at specific radii in the simulation, as
    indicated by the legend.}
    \label{fig:beaming}
\end{figure}

\section{Beamed emission}
\label{postproc}

The most important quantity, which is the 
signature of all ULXs, is a large apparent isotropic luminosity,
that is, $L_\mathrm{iso}=4\pi d^2 F$, measured
from an observed flux, $F$, emitted by an object at a distance, $d$, away 
from the observer (neglecting cosmological effects).
While it is impossible to predict $L_\mathrm{iso}$ reliably from the simulation 
data without sophisticated radiation post-processing, 
we can estimate it at a few locations in the simulation
and see how it changes with radius. Plotted in the
left panel of Fig.~\ref{fig:beaming} is $L_\mathrm{iso}$ as a function of viewing angle for 
different radii. The figure is presented in polar 
coordinates to emphasize the beaming pattern. One
can immediately see that the vast majority of the 
flux is confined within $20^\circ$ of the axis. The peak
$L_\mathrm{iso}$ lies along the axis, and while it decays
with radius (from the continuous green line through the dashed lines to the dotted one), it appears to be converging to a large value
well above $100\,\Ledd$ which is about 
$10^{40}\,\text{erg s}^{-1}$ for a canonical $1.4\,M_\odot$
neutron star.

The apparent luminosity is clearly bright enough for the neutron star to 
qualify as a ULX. To compare the measured degree of beaming with the KLK model, we need to choose a quantity to function as the total luminosity. We consider $L_r$ and $L_{Be}$ as
lower and upper limits respectively 
(although we are confident that the true luminosity is much
closer to $L_r$). These also correspond to lower and upper
limits for the beaming factor $b=L/L_\mathrm{iso}$.

Near the axis, $L_\mathrm{iso}/L_{Be} \sim 25$, and is more or less constant with radius. This already exceeds $1/b$ as computed from the KLK
model \citep{king+17, king+19, king+20} for all the sources they included. 
If we instead consider 
$L_\mathrm{iso}/L_r$, then $1/b$ shoots up to above 80. It
falls at larger radii, to around 60.\footnote{As $r$ increases $\tau_r$ feels the effect of a finite outer boundary, so it is possible that the computed value of $L_r$ may be slightly overestimated, leading to a slightly underestimated $1/b$ at large radii.} 

Note that the beaming factor is a function of the angle. 
In general, $1/b$ is proportional
to flux, which is a function
of $\theta$, and lower viewing angles will tend to display more extreme beaming 
(corresponding to smaller values of $b$). 

Regardless of the measure used to compute the total luminosity, we have
shown that a low-magnetic-field neutron star can produce emission
that is sufficiently beamed to produce a ULX.
The closest theoretical model to our simulation, the KLK model, also predicts large amounts of beaming, although
not as extreme as we observe.
The two models need not completely agree
since they differ in several ways.
The disk in our simulation has a very
large spherization radius when compared to the magnetospheric
radius. The KLK model requires that these two
radii be much closer. This would substantially affect the
outflows, as they are only launched below
$r_\mathrm{sph}$ and above $r_\mathrm{M}$. Including 
the spherization radius in the domain is challenging,
firstly, because it takes a long time for the simulation to 
converge to such large radii, and secondly, because beyond
$r_\mathrm{sph}$, the disk should be similar to a
thin disk, and thin disks are notoriously difficult to simulate.

One issue, although one that we already have plans to remedy, is the limitations of {\koral} when simulating
the radiation field. {\koral} is a grey code which
uses the $M_1$ closure scheme to transport radiation. 
$M_1$ works well for large extended sources, but when 
radiation originating from two or more distinct locations collide, 
the beams interact. 
For us, this is most problematic in the region directly outside
of the accretion columns. The radiation flows towards the
axis and is then directed upward due to the polar boundary condition which is reflective.
In reality, we expect the beams from the accretion
columns to scatter off of the opposite wall formed by the gas
outflows. After enough scatterings, the radiation should be 
collimated and largely moving along the axis.
From Fig.~\ref{fig:rhorad} we can see that the radiation
already appears collimated as soon as it leaves the accretion
column. 

To get a more accurate representation of the
radiation field, we need to go beyond the $M_1$ scheme. 
\texttt{HEROIC} \citep{heroic} 
is a post-processing radiative transfer code that
could be used to recompute the radiation field and
spectrum for an observer at infinity as a
function of viewing angle. While such calculations
are out of the scope of this letter, we plan to 
apply \texttt{HEROIC} post-processing analysis 
to the rest of our simulations in a future
publication.

\section{Summary and Conclusions}
\label{conclusion}

We performed a 2D axisymmetric radiative GRMHD simulation
of accretion onto a neutron star with a 
$2\times 10^{10}\,\text{G}$ dipolar magnetic 
field. The combination of the hard surface
and confinement of the gas into accretion columns by the
stellar magnetic field
near the stellar surface allows the flow to 
release radiative energy at a rate of several
times the Eddington limit. The fraction of this
energy which is able to reach the observer, as
opposed to being absorbed by the outflows, is
difficult to calculate, but a lower limit of the observable
luminosity
should correspond to about $2.5 \, \Ledd$. 
The radiation easily escapes into the polar region 
which is largely devoid of gas due to a combination 
of the magnetic field and rotation
of the outflowing gas which collimates
the radiation flow. While a more precise calculation 
of the radiation field is required due to the limitations 
of the simulation, our results show that this escaping 
radiation will be highly beamed. 
The apparent isotropic luminosity of the source observed pole-on
should be on the order of $100\,\Ledd$. This is
encouraging if we wish to interpret the accreting system
as a model of a neutron-star-powered ULX. 

When compared to the KLK model \citep{king+17, king+19, king+20}, 
we find that the intensity
of the beaming is larger, although we have reason to
believe that post-processing would show a less intensely
beamed distribution of radiation at infinity. 
Furthermore our simulation does not model the same system as
considered by KLK. 
The distance between the {\alfven}
radius and spherization radius is large.
We hope to produce simulations in future studies which can 
reproduce additional observable features of ULXs and which can provide more
information about the nature of the magnetic field in 
pulsating and non-pulsating ULXs.

\acknowledgments
DA thanks Aleksander Sądowski and Andrew Chael for guidance on working 
with the {\koral} code, Maciek Wielgus, Miljenko Cemeljic, 
Jean-Pierre Lasota, and Alexander Tchekhovskoy for their advice, suggestions,
and useful conversations, and Katarzyna Rusinek-Abarca for support, companionship, and patience. 
DA was supported in part by Polish National Science for Center Preludium grant 2017/27/N/ST9/00992. Research supported in part by the 
Polish National Center for Science grant 2019/33/B/ST9/01564.
Computations in this work were carried out on the
Cyfronet Prometheus cluster, part of the PLGrid computing network.

%

\vspace{5mm}
\facility{Cyfronet Prometheus}


\software{Koral, Numpy, SciPy, Matplotlib, Pandas}

\vspace{12cm}

\bibliography{boundary}{}
\bibliographystyle{aasjournal}



\end{document}